\documentclass[11pt]{article}
\usepackage[a4paper,scale=0.8]{geometry}
\usepackage{amsmath}
\usepackage{amsthm}
\usepackage{amssymb}
\usepackage{bm}  % Must be after txfonts
\usepackage{url}
\usepackage[numbers]{natbib}
\usepackage{enumerate}
\usepackage{graphicx}

\newcommand{\vct}{\bm}
\newcommand{\X}{{\mathrm{X}}}
\newcommand{\A}{{\mathrm{A}}}
\newcommand{\B}{{\mathrm{B}}}
\newcommand{\QB}{{\mathrm{QB}}}
\newcommand{\HDB}{{\mathrm{HDB}}}
\newcommand{\K}{{\mathrm{K}}}
\newcommand{\RR}{{\mathbb{R}}}
\newcommand{\trans}{{\mathrm{T}}}
\newcommand{\CUT}{{\mathrm{CUT}}}
\newcommand{\CutPpm}{\mathrm{Cut}}
\newcommand{\CorP}{\mathrm{COR}^\square}
\newcommand{\RMetPpm}{\mathrm{RMet}}
\newcommand{\RCMetP}{\mathrm{RCMet}}
\newcommand*{\abs}[1]{\lvert#1\rvert}
\newcommand*{\avg}[1]{\langle#1\rangle}
\newcommand{\roteq}{{\rotatebox[origin=c]{90}{$=$}}}
\newcommand{\rotsubseteq}{{\rotatebox[origin=c]{90}{$\subseteq$}}}
\newcommand{\rotsubsetneq}{{\rotatebox[origin=c]{90}{$\subsetneq$}}}

\newcommand{\finite}{{\mathrm{finite}}}

\newcommand{\calE}{{\mathcal{E}}}
\newcommand{\calH}{{\mathcal{H}}}
\newcommand{\calQ}{{\mathcal{Q}}}
\newcommand{\QCut}{{\mathcal{Q}_{\mathrm{Cut}}}}

\newcommand{\Fcf}{Correlation}
\newcommand{\fcf}{correlation}

\DeclareMathOperator{\tr}{tr}

\title{On the Relationship between Convex Bodies \\
  Related to Correlation Experiments with Dichotomic Observables}
\author{David Avis$^1$, Hiroshi Imai$^{2,3}$ and Tsuyoshi Ito$^{2,4}$ \\
  \small
  \texttt{avis@cs.mcgill.ca},
  \texttt{imai@is.s.u-tokyo.ac.jp},
  \texttt{tsuyoshi@is.s.u-tokyo.ac.jp} \\
  \small
  \begin{tabular}{cl}
    1 & School of Computer Science, McGill University, \\
      & 3480 University, Montreal, Quebec, Canada H3A 2A7 \\
    2 & Department of Computer Science,
        Graduate School of Information Science and Technology, \\
      & The University of Tokyo, \\
      & 7-3-1 Hongo, Bunkyo-ku, Tokyo 113-0033, Japan \\
    3 & ERATO-SORST Quantum Computation and Information Project, \\
      & Japan Science and Technology Agency, \\
      & 5-28-3 Hongo, Bunkyo-ku, Tokyo 113-0033, Japan \\
    4 & Japan Society for the Promotion of Science
  \end{tabular}
}
\date{May 17, 2006}

\theoremstyle{plain}
\newtheorem{theorem}{Theorem}
\newtheorem{lemma}{Lemma}
\newtheorem{proposition}{Proposition}
\newtheorem{corollary}{Corollary}
\theoremstyle{remark}
\newtheorem{remark}{Remark}

\let\cite\citep

\begin{document}
\maketitle

\begin{abstract}
In this paper we explore further the connections between
convex bodies related to quantum correlation experiments
with dichotomic variables
and related bodies studied in combinatorial optimization,
especially cut polyhedra.
Such a relationship was
established in Avis, Imai, Ito and Sasaki
(2005 \emph{J.\ Phys.\ A: Math.\ Gen.}\ \textbf{38} 10971--87)
with respect to Bell inequalities. We show that several well known
bodies related to cut polyhedra are equivalent to bodies such
as those
defined by Tsirelson (1993 \emph{Hadronic J.\ S.}\ \textbf{8} 329--45)
to represent
hidden deterministic behaviors, quantum behaviors,
and no-signalling behaviors.
Among other things, our results
allow a unique representation of these bodies, give a necessary
condition for vertices of the no-signalling polytope, and give a
method
for bounding the quantum violation of Bell inequalities
by means of a body that contains the set of quantum behaviors.
Optimization over this latter
body may be performed efficiently by semidefinite programming.
In the second part of the paper we apply
these results to the study of classical correlation functions.
We provide a complete list of tight inequalities for the two party
case with $(m,n)$ dichotomic observables when $m=4,n=4$ and
when $\min\{m,n\}\le3$, and give a new general family of
correlation inequalities.
\end{abstract}

\noindent
PACS classification numbers: 03.65.Ud, 02.40.Ft, 02.10.Ud


\section{The convex sets arising from the classical, quantum
  and no-signaling correlation experiments}  \label{sect:sets}

Our terminology for bodies related to quantum correlations follows
closely that of Tsirelson~\cite{Tsi-HJS93} whilst for cut polyhedra
we follow that of Deza and Laurent~\cite{DezLau:cut97}.
We consider the following \emph{(quantum) correlation experiment}.
Suppose that two parties, say Alice and Bob, share a quantum mixed
state $\rho$, or a nonnegative Hermitian operator $\rho$ on a Hilbert
space $\calH_\A\otimes\calH_\B$ with $\tr\rho=1$.
Here $\calH_\A$ and $\calH_\B$ are Hilbert spaces representing the
subsystems owned by Alice and Bob, respectively.
Alice has $m$ $\pm1$-valued observables $A_1,\dots,A_m$ in the space
$\calH_\A$, i.e.\ Hermitian operators $A_i$ on $\calH_\A$ whose
eigenvalues are within $[-1,1]$.
Similarly Bob has $n$ $\pm1$-valued observables $B_1,\dots,B_n$ in
$\calH_\B$.
Alice and Bob measure one observable each, say $A_i$ and $B_j$.
By repeating this process with different choices of $i,j$, we collect
the probabilities $q_{ab\mid ij}$ with which $A_i$ measures to
$a$ and
$B_j$ to $b$ simultaneously under the condition that Alice measures
$A_i$ and Bob $B_j$ for $1\le i\le m$, $1\le j\le n$,
$a,b\in\{\pm1\}$.
Using $\rho,A_i,B_j$, these probabilities are calculated as
$q_{ab\mid ij}=\tr(\rho(\frac{I+aA_i}{2}\otimes\frac{I+bB_j}{2}))$.
The result of such a correlation experiment can be seen as a
$4mn$-dimensional real vector $\vct{q}\in\RR^{4mn}$.
A correlation experiment is said to be \emph{classical} if the state
$\rho$ is separable.

All possible results of (quantum) correlation experiments satisfy the
following three conditions.
The \emph{nonnegativity condition} is $q_{ab\mid ij}\ge0$ for $1\le i\le m$,
$1\le j\le n$, $a,b\in\{\pm1\}$.
The \emph{normalization condition} is $\sum_{a,b\in\{\pm1\}}q_{ab\mid ij}=1$
for $1\le i\le m$ and $1\le j\le n$.
The \emph{no-signaling condition} means that there exist marginal
probabilities $q^\A_{a\mid i}=\sum_{b\in\{\pm1\}} q_{ab\mid ij}$ and
$q^\B_{b\mid j}=\sum_{a\in\{\pm1\}} q_{ab\mid ij}$ which are independent of
$j$ and $i$, respectively.
A vector $\vct{q}\in\RR^{4mn}$ satisfying these conditions is called a
\emph{behavior}, and a vector which is a possible result of a quantum
(resp.\ classical) correlation experiment is called a
\emph{quantum (resp.\ hidden deterministic) behavior}~\cite{Tsi-HJS93}.
We denote the sets of all behaviors, of all quantum behaviors and of
all hidden deterministic behaviors by $X_\B(m,n)$, $X_\QB(m,n)$ and
$X_\HDB(m,n)$, respectively.
Whereas Tsirelson~\cite{Tsi-HJS93} defines them in the general case where each
observable
has an arbitrary number of outcomes, we consider the special case that each
observable has two possible outcomes.

Froissart~\cite{Fro-NC81} shows that $X_\HDB(m,n)$ is a
$(mn+m+n)$-dimensional convex
polytope which has $2^{m+n}$ vertices corresponding to the cases where
observables $A_i$ and $B_j$ are fixed constant $+1$ or $-1$, or in
other words, $A_1,\dots,A_m,\allowbreak B_1,\dots, B_n\in\{\pm I\}$.

A linear inequality on $q_{ab\mid ij}$ which is satisfied for
all possible results of classical correlation experiments, or for all
the points in $X_\HDB(m,n)$, is called a \emph{Bell inequality}
(for $(m,n)$ settings).
However, this is cumbersome for certain purposes because, as is
pointed out by Froissart~\cite{Fro-NC81}, adding any linear
combination of the normalization and no-signaling conditions to an
inequality gives apparently different representations of essentially
the same inequality.
To avoid this, we consider a full-dimensional polytope isomorphic to
$X_\HDB(m,n)$, which will be described shortly.
The set $X_\QB(m,n)$ is a $(mn+m+n)$-dimensional convex, bounded,
closed set~\cite{Tsi-HJS93}.
Recently, Barnett, Linden, Massar, Pironio, Popescu and
Roberts~\cite{BarLinMasPirPopRob-PRA05} studied the vertices of the
polytope consisting of all behaviors with two observables per party.
We call $X_\B(m,n)$ the \emph{no-signaling polytope} following
\cite{BarLinMasPirPopRob-PRA05}.

Due to the normalization and no-signaling conditions,
a behavior $\vct{q}$ is completely specified by the values of
$p_{\A_i\B_j}=q_{-1,-1\mid ij}$,
the marginal probabilities $p_{\A_i}=q^\A_{-1\mid i}$ and
$p_{\B_j}=q^\B_{-1\mid j}$.
We consider $\vct{p}$ as an $(mn+m+n)$-dimensional vector in the vector
space $\RR^{V_{m,n}\cup E_{m,n}}$,
where $V_{m,n}=\{\A_1,\dots,\A_m,\allowbreak\B_1,\dots,\B_n\}$ and
$E_{m,n}=\{\A_i\B_j\mid 1\le i\le m,\;1\le j\le n\}$ are the node set
and the edge set, respectively, of the complete bipartite graph
$\K_{m,n}$.
We denote the vector $\vct{q}\in\RR^{4mn}$ corresponding to a given vector
$\vct{p}\in\RR^{V_{m,n}\cup E_{m,n}}$ by $\iota(\vct{p})$.
Formally, $\iota$ is a one-to-one affine mapping
from $\RR^{V_{m,n}\cup E_{m,n}}$ into $\RR^{4mn}$ defined by
$q_{-1,-1\mid ij}=p_{\A_i\B_j}$, $q_{-1,+1\mid ij}=p_{\A_i}-p_{\A_i\B_j}$,
$q_{+1,-1\mid ij}=p_{\B_j}-p_{\A_i\B_j}$,
$q_{+1,+1\mid ij}=1-p_{\A_i}-p_{\B_j}+p_{\A_i\B_j}$.

This means that we can consider the convex bodies (convex, bounded,
closed, full-dimensional sets) $\iota^{-1}(X_\B(m,n))$,
$\iota^{-1}(X_\QB(m,n))$ and $\iota^{-1}(X_\HDB(m,n))$ instead of
$X_\B(m,n)$, $X_\QB(m,n)$ and $X_\HDB(m,n)$, respectively.
Especially, $\iota^{-1}(X_\HDB(m,n))$ is exactly identical to the
\emph{correlation polytope}~\cite{Pit:prob89} of the complete
bipartite graph $\K_{m,n}$, which we denote by $\CorP(\K_{m,n})$
following Deza and Laurent~\cite{DezLau:cut97}.
The correlation polytope of a graph is introduced by
Pitowsky~\cite{Pit:prob89} (see also \cite{Pit-MP91})
to describe the possible results
of classical correlation experiments in a broader sense than our use
of the term, and our case corresponds to the correlation polytope of
the complete bipartite graph $\K_{m,n}$.
The correlation polytope has been also studied in context of
combinatorial optimization under the name ``boolean quadric polytope''
in relation to unconstrained quadratic 0-1
programming~\cite{Pad-MP89} (see Section~5.1 of \cite{DezLau:cut97}).
We denote $\iota^{-1}(X_\QB(m,n))$ by $\calQ(m,n)$.
We refer to $\calQ(m,n)$ as the \emph{quantum correlation set}.

Bell inequalities can be written by using
$p_{\A_i},p_{\B_j},p_{\A_i\B_j}$ instead of $q_{ab\mid ij}$.
Using vectors $\vct{p}$, a Bell inequality is a linear inequality
satisfied for all the points in $\CorP(\K_{m,n})$.
This avoids the problem stated above because $\CorP(\K_{m,n})$ is
full-dimensional and representation of an inequality is unique up to
positive scaling.
From now on, we represent Bell inequalities in terms of $\vct{p}$.

Aside from trivial examples such as $p_{\A_1\B_1}\ge0$,
$p_{\A_1\B_1}-p_{\B_1}\le0$ or $p_{\A_1}+p_{\B_1}-p_{\A_1\B_1}\le1$,
a nontrivial example with two observables per party is the famous
Clauser-Horne-Shimony-Holt (CHSH)
inequality~\cite{ClaHorShiHol-PRL69}:
\begin{equation}
  -p_{\A_1}-p_{\B_1}
  +p_{\A_1\B_1}+p_{\A_1\B_2}+p_{\A_2\B_1}-p_{\A_2\B_2}\le0.
  \label{eq:chsh-cor}
\end{equation}
Bell inequalities are exactly linear inequalities valid for the correlation
polytope $\CorP(\K_{m,n})$.
Any inequality that can be described as a sum of two different Bell
inequalities is trivially a Bell inequality, and
such a Bell inequality is said to be \emph{redundant}.
A Bell inequality is said to be \emph{tight} if not redundant.
From the central theorem in the theory of convex polytopes (see e.g.\
Section~1.1 of \cite{Zie:lectures98}), for any fixed $m$ and $n$,
there are finitely many tight Bell inequalities, which give a unique minimum
representation of the correlation polytope $\CorP(\K_{m,n})$ by
inequalities.

The most famous example of a linear inequality valid for the quantum
correlation set is Tsirelson's inequality~\cite{Cir-LMP80} stating that
the maximum violation of the CHSH inequality~(\ref{eq:chsh-cor}) in the
quantum case is $\sqrt{2}-1$.
This maximum is achieved by using a maximally entangled pure state
in the two-qubit system and suitable observables.
Pitowsky~\cite{Pit-QTRF02} also considers this set.
We note that Tsirelson~\cite{Cir-LMP80} states an exact
characterization of the set $X_\QB(2,2)$ by a system of algebraic
equations and inequalities on finitely many variables with
quantifiers.

\begin{figure}
  \[
    \begin{array}{ccccccccc}
      & \RR^{4mn}
        & \stackrel{\iota}{\leftarrow}
          & \RR^{V_{m,n}\cup E_{m,n}}
            & \stackrel{\varphi}{\to}
              & \RR^{\nabla E_{m,n}}
                & \stackrel{\text{\makebox[0pt]{$\pi$ (see Prop.~\ref{prop:cf-kmn})}}}{\to}
                  & \RR^{E_{m,n}}
                    & \text{Complexity} \\
      & \rotsubseteq
        &
          & \rotsubseteq
            &
              & \rotsubseteq
                &
                  & \rotsubseteq
                    & \\
      \text{No-signaling}
      & X_\B(m,n)
        & \stackrel{\text{Th.~\ref{theorem:nosig-rcmet}}}{\cong}
          & \RCMetP(\K_{m,n})
            & \cong
              & \RMetPpm(\nabla\K_{m,n})
                & \stackrel{\text{Prop.~\ref{prop:rmet-cube}}}{\to}
                  & [-1,1]^{E_{m,n}}
                    & \text{Easy} \\
      &
        &
          &
            &
              & \rotsubsetneq
                &
                  & \rotsubsetneq
                    & \\
      & \rotsubsetneq
        &
          & \rotsubsetneq
            &
              & \makebox[0pt]{$\calE(\nabla\K_{m,n})\cap\RMetPpm(\nabla\K_{m,n})$}\;\;\;\;\;\;\;
                & \to
                  & \calE(\K_{m,n})
                    & \text{Tractable} \\
      &
        &
          &
            &
              & \rotsubseteq\rlap{\footnotesize\ Th.~\ref{theorem:quantum-elliptope2}}
                &
                  & \roteq\rlap{\footnotesize\ Cor.~\ref{cor:quantum-elliptope}}
                    & \\
      \text{Quantum}
      & X_\QB(m,n)
        & \cong
          & \calQ(m,n)
            & \cong
              & \QCut(m,n)
                & \to
                  & M_\QB(m,n)
                    & \text{Unknown\footnotemark} \\
      & \rotsubsetneq
        &
          & \rotsubsetneq
            &
              & \rotsubsetneq
                &
                  & \rotsubsetneq
                    & \\
      \text{Classical}
      & X_\HDB(m,n)
        & \cong
          & \CorP(\K_{m,n})
            & \cong
              & \CutPpm(\nabla\K_{m,n})
                & \to
                  & \CutPpm(\K_{m,n})
                    & \text{Hard}
    \end{array}
  \]
  \caption{Relationship among various convex sets discussed in the
    paper.
    The injective mapping $\iota$ is defined in
    Section~\ref{sect:sets}.
    The isomorphism $\varphi$ is the covariance mapping
    stated in Section~\ref{sect:covariance}.
    The projection $\pi$ is the standard projection from
    $\RR^{\nabla E_{m,n}}$ to $\RR^{E_{m,n}}$.
    Proper inclusions are valid for $m,n\ge2$.%
  }
  \label{fig:overview}
\end{figure}
\footnotetext{Except for $M_\QB(m,n)$, which is equal to
  $\calE(\K_{m,n})$ by Corollary~\ref{cor:quantum-elliptope}.}

Figure~\ref{fig:overview} gives an overview of most of the results we
will see in this paper.
The two leftmost columns labeled as ``$\RR^{4mn}$'' and
``$\RR^{V_{m,n}\cup E_{m,n}}$'' depict the relationship explained so
far except for the relations involving $\RCMetP(\K_{m,n})$,
the rooted correlation semimetric polytope introduced in the following
section.
The ``complexity'' column refers to the computational complexity of
testing membership in the given body.
The rest of the figure will be explained in the following sections.

\section{The no-signaling polytope and the rooted correlation semimetric
  polytope}

We will prove that the no-signaling polytope, if represented in terms
of vectors $\vct{p}\in\RR^{V_{m,n}\cup E_{m,n}}$ instead of the
vectors $\vct{q}\in\RR^{4mn}$, is identical to a convex polytope
which arises in the combinatorial optimization.
The \emph{rooted correlation semimetric polytope} of a graph $G=(V,E)$
is the convex polytope $\RCMetP(G)$ in $\RR^{V\cup E}$
defined by a system of inequalities
$p_{uv}\ge0$, $p_{u}-p_{uv}\ge0$, $1-p_{u}-p_{v}+p_{uv}\ge0$
for $uv\in E$~\cite{Pad-MP89}.
Padberg~\cite{Pad-MP89} studies this polytope as a natural linear
relaxation of the correlation polytope and investigates the
relationship between them.
The name ``rooted correlation semimetric polytope,'' used in Deza and
Laurent~\cite{DezLau:cut97},
comes from its relation to the cut polytope explained in the next
section.
We note that $\RCMetP(\K_n)$ appears in Pitowsky~\cite{Pit-JMP86}.

\begin{theorem} \label{theorem:nosig-rcmet}
  The no-signaling polytope $X_\B(m,n)$ satisfies
  $X_\B(m,n)=\iota(\RCMetP(\K_{m,n}))$,
  where $\RCMetP(\K_{m,n})$ denotes the rooted correlation semimetric
  polytope of the complete bipartite graph $\K_{m,n}$.
\end{theorem}

\begin{proof}
  Since every point in $X_\B(m,n)$ satisfies the normalization and
  the no-signaling conditions, the set $X_\B(m,n)$ is contained in
  the image $\iota(\RR^{V_{m,n}\cup E_{m,n}})$.
  Let $\vct{p}\in\RR^{V_{m,n}\cup E_{m,n}}$ and $\vct{q}=\iota(\vct{p})$.
  From the definition of $\iota$,
  the following logical equivalences hold:
  $q_{-1,-1\mid ij}\ge0\iff p_{\A_i\B_j}\ge0$,
  $q_{-1,+1\mid ij}\ge0\iff p_{\A_i}-p_{\A_i\B_j}\ge0$,
  $q_{+1,-1\mid ij}\ge0\iff p_{\B_j}-p_{\A_i\B_j}\ge0$,
  $q_{+1,+1\mid ij}\ge0\iff 1-p_{\A_i}-p_{\B_j}+p_{\A_i\B_j}\ge0$.
  This means that
  $\vct{q}\in X_\B(m,n)$ if and only if $\vct{p}\in\RCMetP(\K_{m,n})$,
  which implies $X_\B(m,n)=\iota(\RCMetP(\K_{m,n}))$.
\end{proof}

The vertices of the rooted correlation semimetric polytope are studied
by Padberg~\cite{Pad-MP89}.

\begin{theorem}[\cite{Pad-MP89}] \label{theorem:rcmet-half-int}
  The coordinates of the vertices of the rooted correlation semimetric
  polytope $\RCMetP(G)$ are in $\{0,1/2,1\}$.
\end{theorem}

\begin{corollary}
  The coordinates of the vertices of $X_\B(m,n)$ are in
  $\{0,1/2,1\}$.
\end{corollary}

\begin{proof}
  Immediate from Theorems~\ref{theorem:nosig-rcmet} and
  \ref{theorem:rcmet-half-int} and the definition of $\iota$.
\end{proof}

In a related work~\cite{BarLinMasPirPopRob-PRA05},
Barrett, Linden, Massar, Pironio, Popescu and
Roberts investigate the vertices of
the no-signaling polytope with two $k$-outcome observables per party.

\section{The covariance mapping and the cut polytope}  \label{sect:covariance}

The correlation polytope $\CorP(\K_{m,n})$ is isomorphic to the
cut polytope of a certain graph, which is the suspension graph of $K_{m,n}$
and has a natural physical interpretation.
The \emph{suspension graph} $\nabla\K_{m,n}$ of $\K_{m,n}$ is obtained
from $\K_{m,n}$ by adding one new node $\X$ which is adjacent to all
the other $m+n$ nodes.
The graph $\nabla\K_{m,n}=(\nabla V_{m,n},\nabla E_{m,n})$ has $1+m+n$ nodes
$\nabla V_{m,n}=\{\X,\A_1,\dots,\A_m,\allowbreak\B_1,\dots,\B_n\}$ and
$m+n+mn$ edges
$\nabla E_{m,n}=\{\X\A_i\mid 1\le i\le m\}\cup
           \{\X\B_j\mid 1\le j\le n\}\cup
           \{\A_i\B_j\mid 1\le i\le m,\;1\le j\le n\}$.
Here we denote the node set and the edge set of $\nabla\K_{m,n}$ by
$\nabla V_{m,n}$ and $\nabla E_{m,n}$, slightly abusing the notation.
For any vector $\vct{c}\in\{\pm1\}^{\nabla V_{m,n}}$, the \emph{cut vector}
of $\nabla\K_{m,n}$ defined by $\vct{c}$ is the vector
$\vct{x}\in\RR^{\nabla E_{m,n}}$ such that $x_{uv}=c_u c_v$ for
$uv\in \nabla E_{m,n}$.
The convex hull of all the cut vectors of $\nabla\K_{m,n}$ is called the
\emph{cut polytope} of $\nabla\K_{m,n}$ and denoted by
$\CutPpm(\nabla\K_{m,n})$.
Research on the cut polytope has long and rich history and many results
are known,
see Deza and Laurent~\cite{DezLau:cut97}.
We note that in \cite{DezLau:cut97} and many other papers in
combinatorics, the cut polytope is defined in the
terms of 0/1 cut vectors instead of $\pm1$ cut vectors,
which we use here for better correspondence to the $\pm1$-valued
observables.
All the results on the cut polytope can be stated both in the $\pm1$
terminology and in the 0/1 terminology.

The correlation polytope and the cut polytope are related via
the covariance mapping~\cite[Section~5.2]{DezLau:cut97}.
The correlation polytope $\CorP(\K_{m,n})$ is isomorphic to the cut
polytope $\CutPpm(\nabla\K_{m,n})$ via a linear isomorphism, called the
\emph{covariance mapping} $\varphi$, which maps
$\vct{p}\in\RR^{V_{m,n}\cup E_{m,n}}$ to $\vct{x}\in\RR^{\nabla E_{m,n}}$
defined by $x_{\X\A_i}=1-2p_{\A_i}$, $x_{\X\B_j}=1-2p_{\B_j}$ and
$x_{\A_i\B_j}=1-2p_{\A_i}-2p_{\B_j}+4p_{\A_i\B_j}$.

In the classical and quantum cases,
the coordinates of the vector $\vct{x}$ have a natural physical
interpretation
related to the observables $A_1,\dots,A_m,B_1,\dots,B_n$
as stated in the following proposition.

\begin{proposition} \label{prop:cut-expect}
  Let a vector $\vct{p}\in\RR^{V_{m,n}\cup E_{m,n}}$ be as defined
  in Section~\ref{sect:sets}, and let $\vct{x}=\varphi(\vct{p})$.
  Then $x_{\X\A_i}=\avg{A_i}$, $x_{\X\B_j}=\avg{B_j}$
  and $x_{\A_i\B_j}=\avg{A_iB_j}$ for $1\le i\le m$ and $1\le j\le n$,
  where $\avg{{\cdot}}$ denotes the expected value.
\end{proposition}

\begin{proof}
  The equation $x_{\X\A_i}=\avg{A_i}$ is proved as follows:
  $\avg{A_i}=(+1)\cdot(1-p_{\A_i})+(-1)\cdot p_{\A_i}=1-2p_{\A_i}
            =x_{\X\A_i}$.
  The equations $\avg{B_j}=x_{\X\B_j}$ and $\avg{A_iB_j}=x_{\A_i\B_j}$
  can be verified similarly.
\end{proof}

The image of $\RCMetP(\K_{m,n})$ under the covariance mapping $\varphi$
is the
\emph{rooted semimetric polytope} of $\nabla\K_{m,n}$ pointed at the node
$\X$~\cite[Section~27.2]{DezLau:cut97},
and is denoted by $\RMetPpm(\nabla\K_{m,n})$.

\begin{proposition} \label{prop:nosig-rmet}
  The polytope $\RMetPpm(\nabla\K_{m,n})=
  \varphi(\RCMetP(\K_{m,n}))$ is defined by
  inequalities $x_{\X\A_i}+x_{\X\B_j}+x_{\A_i\B_j}\ge-1$,
  $-x_{\X\A_i}-x_{\X\B_j}+x_{\A_i\B_j}\ge-1$,
  $x_{\X\A_i}-x_{\X\B_j}-x_{\A_i\B_j}\ge-1$ and
  $-x_{\X\A_i}+x_{\X\B_j}-x_{\A_i\B_j}\ge-1$
  for $1\le i\le m$, $1\le j\le n$.
\end{proposition}

We denote the image of $\calQ(m,n)$ under the covariance mapping
$\varphi$ by $\QCut(m,n)=\varphi(\calQ(m,n))$.

\section{Correlation functions}
\label{sect:cor-func}

The \emph{correlation function} $x_{\A_i\B_j}$ is the expected value
$\avg{A_iB_j}$ of the product of an observable by Alice and another
observable by Bob.
The correlation functions $x_{\A_i\B_j}$ for all $i,j$ form an
$mn$-dimensional vector $\vct{x}'\in\RR^{E_{m,n}}$.
Tsirelson~\cite{Cir-LMP80,Tsi-JSovM87,Tsi-HJS93} gives a detailed
study on the sets of correlation functions which are possible in
classical and quantum correlation experiments.

Clearly a correlation function $x_{\A_i\B_j}$ takes a value in the
range $[-1,1]$.
The value $(1-x_{\A_i\B_j})/2$ in the range $[0,1]$ is the probability
of the exclusive OR of the events that $A_i$ measures to $-1$ and that
$B_j$ to $-1$ under the condition that Alice measures $A_i$ and Bob
$B_j$.
We note that if we replace the ``exclusive OR'' by ``AND,'' we obtain
the probability with which both $A_i$ and $B_j$ measure to $-1$ under
the same condition, resulting in another convex polytope explored by
Froissart~\cite{Fro-NC81}.

From Proposition~\ref{prop:cut-expect}, the vector $\vct{x}'$ is the
projection of $\vct{x}$ defined in Section~\ref{sect:covariance} to
$\RR^{E_{m,n}}$ by the standard projection
$\pi\colon\RR^{\nabla E_{m,n}}\to\RR^{E_{m,n}}$,
giving the following proposition.

\begin{proposition} \label{prop:cf-kmn}
  Let $\pi\colon\RR^{\nabla E_{m,n}}\to\RR^{E_{m,n}}$ be the standard
  projection.
  \begin{enumerate}[(i)]
  \item
    The vectors of correlation functions which are possible in
    classical correlation experiments form the cut polytope
    $\CutPpm(\K_{m,n})$ of the complete bipartite graph $\K_{m,n}$.
    Here the cut polytope $\CutPpm(\K_{m,n})$ is a convex polytope in
    $\RR^{E_{m,n}}$ defined in the same way as $\CutPpm(\nabla\K_{m,n})$,
    replacing $\nabla E_{m,n}$ by $E_{m,n}$.
  \item
    The vectors of correlation functions which are possible in
    quantum correlation experiments form a convex body
    $\pi(\QCut(m,n))$.
  \item
    The vectors of correlation functions which can arise from
    correlation tables satisfying the nonnegativity, normalization and
    no-signaling conditions form a convex polytope
    $\pi(\RMetPpm(\nabla\K_{m,n}))$.
  \end{enumerate}
\end{proposition}

\begin{proof}
  Immediate from Proposition~\ref{prop:cut-expect}, with
  Theorem~\ref{theorem:nosig-rcmet} for (iii).
\end{proof}

Similar to a Bell inequality, a \emph{\fcf\ inequality}
(for $(m,n)$ settings) is a linear inequality on correlation
functions which is satisfied for all possible results of classical
correlation experiments.
The CHSH inequality~(\ref{eq:chsh-cor}) can be also written as a
correlation inequality:
\begin{equation}
  x_{\A_1\B_1}+x_{\A_1\B_2}+x_{\A_2\B_1}-x_{\A_2\B_2}\le2.
  \label{eq:chsh-cut}
\end{equation}

An $N$-party version of the set of classical correlation functions
with two observables per party is studied by Werner and
Wolf~\cite{WerWol-PRA01}, and it turns out to be a crosspolytope, a
convex polytope with a surprisingly simple structure compared to the
complicated structure of the correlation polytope.
One might expect that there is also a simple characterization of
the correlation inequalities for the case with an arbitrary number of
observables per party.
In two-party case, this leads to an analysis of a projection of the
correlation polytope $\CorP(\K_{m,n})$ and the cut polytope
$\CutPpm(\nabla\K_{m,n})$, which is the cut polytope $\CutPpm(\K_{m,n})$
of the bipartite graph $\K_{m,n}$.
However as shown in \cite{AviDez-Net91,Pit-MP91}, such a
characterization is unlikely, since membership testing in these polyhedra
is NP-complete.

A \fcf\ inequality for $(m,n)$ settings is said to be
\emph{redundant} if it is a sum of two \fcf\ inequalities
for $(m,n)$ settings which are not the scalings of it, and
\emph{tight} if not redundant.
Tight \fcf\ inequalities are exactly the facet-inducing inequalities of
the corresponding polytope $\CutPpm(\K_{m,n})$.

There are some ``obvious'' symmetries acting on Bell inequalities, and
they act also on \fcf\ inequalities~\cite{WerWol-PRA01}.
They are combinations of one or more of the following basic
symmetries: (i) party exchange, (ii) observable exchange, and (iii)
relabeling of outcomes (these terms are coined by
Masanes~\cite{Mas-QIC03}).
Two \fcf\ inequalities are said to be \emph{equivalent} if one of them
can be transformed to the other by applying one or more of these basic
symmetries.
The cut polytope admits two basic symmetries called permutation and
switching~\cite[Sections~26.2, 26,3]{DezLau:cut97}.
As is the case of Bell inequalities and $\CutPpm(\nabla\K_{m,n})$ discussed in
\cite{AviImaItoSas-JPA05}, the basic symmetries acting on \fcf\
inequalities correspond exactly to the basic symmetries of
$\CutPpm(\K_{m,n})$.

In the no-signaling case, the corresponding polytope becomes the hypercube.

\begin{proposition}
  \label{prop:rmet-cube}
  The set $\pi(\RMetPpm(\nabla\K_{m,n}))$, which is the set of vectors
  in $\RR^{E_{m,n}}$ realizable as the correlation function arising
  from no-signaling correlation tables, is equal to the hypercube
  $[-1,1]^{E_{m,n}}$.
\end{proposition}

\begin{proof}
  It is trivial that
  $\pi(\RMetPpm(\nabla\K_{m,n}))\subseteq[-1,1]^{E_{m,n}}$.

  To prove the converse, let $\vct{x}'\in[-1,1]^{E_{m,n}}$.
  We define $\vct{x}\in\RR^{\nabla E_{m,n}}$ by $x_{\A_i\B_j}=x'_{\A_i\B_j}$
  and $x_{\X\A_i}=x_{\X\B_j}=0$ for $1\le i\le m$ and $1\le j\le n$.
  Then $\vct{x}\in\RMetPpm(\nabla\K_{m,n})$ and $\pi(\vct{x})=\vct{x}'$,
  which implies
  $\pi(\RMetPpm(\nabla\K_{m,n}))\supseteq[-1,1]^{E_{m,n}}$.
\end{proof}

The implication of this theorem is that all correlations between
observables $A_i,B_j$ are possible under the no-signaling condition
alone.

In the quantum case, Tsirelson's theorem~\cite{Cir-LMP80}
(see \cite{Tsi-JSovM87} for a proof)
gives a beautiful characterization of the quantum bound of correlation
functions.

\begin{theorem}[\cite{Tsi-JSovM87}]
  \label{theorem:tsirelson}
  Let $m,n$ be positive integers.
  On a real vector $\vct{x}\in\RR^{E_{m,n}}$, the following
  assertions are equivalent.
  \begin{enumerate}[(i)]
  \item
    $\vct{x}\in\QCut(m,n)$.
    In other words, there exist Hilbert spaces $\calH_\A$ and
    $\calH_\B$, a mixed state $\rho$ on $\calH_\A\otimes\calH_\B$
    and Hermitian operators $A_1,\dots,A_m$ on $\calH_\A$
    and $B_1,\dots,B_n$ on $\calH_\B$ with eigenvalues in $[-1,1]$,
    such that $x_{\A_i\B_j}=\tr[\rho(A_i\otimes B_j)]$
    for $1\le i\le m$, $1\le j\le n$.
  \item
    The same as (i) with the following additional conditions:
    (a) $\calH_\A$ and $\calH_\B$ have finite dimensions $d_\A$ and
        $d_\B$, respectively, and $d_\A\le2^{\lceil m/2\rceil}$,
        $d_\B\le2^{\lceil n/2\rceil}$.
    (b) $A_i^2=B_j^2=I$ and
        $\tr[\rho(A_i\otimes I)]=\tr[\rho(I\otimes B_j)]=0$
        for $1\le i\le m$, $1\le j\le n$.
    (c) Anticommutators $A_{i_1}A_{i_2}+A_{i_2}A_{i_1}$
        for $1\le i_1<i_2\le m$
        and $B_{j_1}B_{j_2}+B_{j_2}B_{j_1}$ for $1\le j_1<j_2\le n$
        are scalar, that is, proportional to $I$.
  \item
    There exist $m+n$ unit vectors
    $\vct{u}_1,\dots,\vct{u}_m,\allowbreak\vct{v}_1,\dots,\vct{v}_n$
    in the vector space $\RR^{m+n}$ such that
    $x_{\A_i\B_j}=\vct{u}_i\cdot\vct{v}_j$.
  \end{enumerate}
\end{theorem}

In combinatorial optimization, vectors whose elements are defined as
inner products of unit vectors, as in condition~(iii), are well studied.
They form a set called the
\emph{elliptope}~\cite[Section~26.4]{DezLau:cut97}.
The elliptope $\calE(G)$ of a graph $G=(V,E)$ with $n=\abs{V}$ nodes
is the convex body consisting of vectors $\vct{x}\in\RR^E$ such that
there exist a unit vector $\vct{u}_i$ in $\RR^n$ for each node
$i\in V$ satisfying $x_{ij}=\vct{u}_i\cdot\vct{u}_j$.
In particular, the elliptope $\calE(\K_{m,n})$ of the complete
bipartite graph $\K_{m,n}$ is the set of the vector
$\vct{x}\in\RR^{E_{m,n}}$ satisfying the condition~(iii) of
Theorem~\ref{theorem:tsirelson}.

\begin{corollary}
  \label{cor:quantum-elliptope}
  $\pi(\QCut(m,n))=\calE(\K_{m,n})$.
\end{corollary}

It is well-known that the elliptope can also be characterized by using
nonnegative definite matrices called Gram matrices (see
e.g.\ Section~28.4.1 of \cite{DezLau:cut97}).
The following theorem is the characterization of $\calE(G)$ in
this form.

\begin{theorem}
  Let $G=(V,E)$ be a graph with $\abs{V}=n$ nodes labeled as
  $1,\dots,n$.
  A vector $\vct{x}\in\RR^E$ satisfies
  $\vct{x}\in\calE(G)$ if and only if there is an
  $n\times n$ real symmetric nonnegative definite matrix
  $H=(h_{ij})$ such that $h_{ij}=x_{ij}$ for all $ij\in E$ and
  $h_{ii}=1$ for all $1\le i\le n$.
\end{theorem}

Linear functions can be optimized efficiently over the elliptope $\calE(G)$ by
using semidefinite programming (see e.g.\ \cite{BoyVan:convex04};
this optimization is the heart of
the Goemans-Williamson approximation algorithm for the maximum cut
problem~\cite{GoeWil-JACM95}).
By efficiently we mean up to some error $\varepsilon$ in time
polynomial in the input size and $log(1/\varepsilon)$.
A similar statement is true about membership testing in $\calE(G)$.
This fact combined with Corollary~\ref{cor:quantum-elliptope} implies
that the maximum violation of given correlation inequalities in the
quantum case can be computed efficiently, as pointed out by Cleve,
H{\o}yer, Toner and Watrous~\cite{CleHoyTonWat-CCC04}.
In addition,
Grothendieck's inequality~\cite{Kri-AdvMath79}, stating that the
elliptope is not much
larger than the cut polytope for bipartite graphs,
gives an upper bound of the violation of
correlation inequalities~\cite{Tsi-JSovM87,CleHoyTonWat-CCC04}.
Besides, as is pointed out by Tsirelson~\cite{Tsi-HJS93},
Grothendieck~\cite{Gro-BSMSP53} proves that for a vector
$\vct{x}\in\RR^{E_{m,n}}$ to belong to $\calE(\K_{m,n})$, it is
necessary that the vector $\vct{y}\in\RR^{E_{m,n}}$ defined by
$y_{\A_i\B_j}=(2/\pi)\arcsin x_{\A_i\B_j}$ belongs to
$\CutPpm(\K_{m,n})$.
Tsirelson conjectures there that this condition is also sufficient for
$m=n=2$.
This condition is known under the name
\emph{cut condition}~\cite[Section~31.3.1]{DezLau:cut97} in
combinatorial optimization, and necessary for a vector to belong to
$\calE(G)$ with any graph $G$ (where the cut condition for $\calE(G)$
is defined analogously).
Laurent~\cite{Lau-LAA97} proves that the cut condition for $\calE(G)$
is sufficient if and only if $G$ has no $\K_4$-minor.
According to this, the cut condition for $\calE(\K_{m,n})$ is sufficient
if and only if $\min\{m,n\}\le2$, and Tsirelson's conjecture is
settled affirmatively.

Pitowsky~\cite{Pit-QTRF02} considers the intersection of
$\calQ(m,n)$ with the subspace $U$ of $\RR^{V_{m,n}\cup E_{m,n}}$
defined by $m+n$ equations $p_{\A_i}=1/2$ and $p_{\B_j}=1/2$,
and proves the following
theorem as a corollary of the equivalence of the conditions~(i) and (ii)
in Theorem~\ref{theorem:tsirelson}.%
\footnote{Strictly speaking, Pitowsky~\cite{Pit-QTRF02} considers the
  subset $\calQ_\finite(m,n)$ of $\calQ(m,n)$ consisting of quantum
  correlation tables which can be realized with $\rho$
  finite-dimensional quantum states.
  \cite{Pit-QTRF02} proves that the affine mapping in the
  Theorem~\ref{theorem:quantum} maps $\calQ_\finite(n,n)$ onto
  $\calQ_\finite(n,n)\cap U$.
  Since Tsirelson's theorem is valid also for infinite-dimensional
  quantum systems, the same proof is valid for infinite-dimensional
  systems.}

\begin{theorem} \label{theorem:quantum}
  An affine mapping from $\RR^{V_{m,n}\cup E_{m,n}}$ to itself
  which maps $\vct{p}$ to $\vct{p}'$ defined by
  $p'_{\A_i}=p'_{\B_j}=1/2$ and
  $p'_{\A_i\B_j}=p_{\A_i\B_j}-\frac12p_{\A_i}-\frac12p_{\B_j}+\frac12$
  maps $\calQ(m,n)$ onto $\calQ(m,n)\cap U$.
\end{theorem}

The affine mapping in Theorem~\ref{theorem:quantum} can be
explained by using the cut polytope and the covariance mapping.
The vector $\varphi(\vct{p}')$ are equal to $\varphi(\vct{p})$ in the
coordinates corresponding to $E_{m,n}$, and zero in the other
coordinates.
Therefore, Theorem~\ref{theorem:quantum} is equivalent to stating
that for any vector $\vct{x}$ in $\varphi(\calQ(m,n))=\QCut(m,n)$,
the vector $\vct{y}$ obtained from $\vct{x}$ by replacing
the $m+n$ coordinates $x_{\X\A_i}$ and $x_{\X\B_j}$ by zero also
belongs to $\QCut(m,n)$.
This property follows from the symmetry of the set
$\QCut(m,n)$, and the same holds also for
$\CutPpm(\nabla\K_{m,n})$ and $\RMetPpm(\nabla\K_{m,n})$.

\section{Implication of Tsirelson's theorem on the quantum correlation set}

In this section we describe a body that contains the quantum correlation set
$\QCut(m,n)$,
and give some related applications. The body
is described in the following theorem, the proof of which is based
on Corollary~\ref{cor:quantum-elliptope} of Tsirelson's theorem.

\begin{theorem}
  \label{theorem:quantum-elliptope2}
  $\QCut(m,n)\subseteq
   \calE(\nabla\K_{m,n})\cap\RMetPpm(\nabla\K_{m,n})$.
\end{theorem}

\begin{proof}
  $\QCut(m,n)=\varphi(\calQ(m,n))\subseteq\RMetPpm(\nabla\K_{m,n})$
  follows from
  $\calQ(m,n)\subseteq X_\B(m,n)$ and
  $\varphi(X_\B(m,n))=\RMetPpm(\nabla\K_{m,n})$.

  To prove $\QCut(m,n)\subseteq\calE(\nabla\K_{m,n})$, let
  $\vct{x}\in\QCut(m,n)$.
  Then there exist Hilbert spaces $\calH_\A$ and $\calH_\B$,
  a quantum state $\rho$ on $\calH_\A\otimes\calH_\B$,
  $m$ Hermitian operators $A_1,\dots,A_m$ on $\calH_\A$ and
  $n$ Hermitian operators $B_1,\dots,B_n$ on $\calH_\B$
  such that $x_{\X\A_i}=\avg{A_i}$, $x_{\X\B_j}=\avg{B_j}$,
  $x_{\A_i\B_j}=\avg{A_iB_j}$.
  We add Hermitian operator $A_{m+1}=I$ and $B_{n+1}=I$ and consider
  the bipartite case $\K_{m+1,n+1}$. Define
  $\vct{y}\in\RR^{E_{m+1,n+1}}$ by $y_{\A_i\B_j}=\avg{A_iB_j}$.
  Then for $1\le i\le m$ and $1\le j\le n$,
  $y_{\A_i\B_j}=x_{\A_i\B_j}$,
  $y_{\A_i\B_{n+1}}=x_{\X\A_i}$, $y_{\A_{m+1}\B_j}=x_{\X\B_j}$ and
  $y_{\A_{m+1}\B_{n+1}}=1$.
  Corollary~\ref{cor:quantum-elliptope} guarantees
  $\vct{y}\in\calE(\K_{m+1,n+1})$.
  This means that there are unit vectors
  $\vct{u}_1,\dots,\vct{u}_{m+1},\allowbreak
   \vct{v}_1,\dots,\vct{v}_{n+1}$ in the vector space
  $\RR^{m+n+2}$ such that $y_{\A_i\B_j}=\vct{u}_i\cdot\vct{v}_j$
  for $1\le i\le m+1$ and $1\le j\le n+1$.
  From $y_{\A_{m+1}\B_{n+1}}=1$, we have $\vct{u}_{m+1}=\vct{v}_{n+1}$.
  Note that the $m+n+2$ vectors $\vct{u}_i$ and $\vct{v}_j$ lie in
  an $(m+n+1)$-dimensional subspace of $\RR^{m+n+2}$ because there are
  at most $m+n+1$ distinct vectors among them.
  This means $\vct{x}\in\calE(\nabla\K_{m,n})$.
\end{proof}

\begin{remark}
  Given Corollary~\ref{cor:quantum-elliptope}, one may expect that
  $\QCut(m,n)=\calE(\nabla\K_{m,n})$, but this can be easily
  disproved as follows.
  Since a vector $\vct{x}\in\RR^{\nabla E_{m,n}}$
  defined by $x_{\X\A_i}=x_{\X\B_j}=x_{\A_i\B_j}=-1/4$
  for all $1\le i\le m$, $1\le j\le n$
  lies in $\calE(\nabla\K_{m,n})\setminus\RMetPpm(\nabla\K_{m,n})$,
  we have $\calE(\nabla\K_{m,n})\nsubseteq\RMetPpm(\nabla\K_{m,n})$
  and therefore $\calE(\nabla\K_{m,n})\nsubseteq\QCut(m,n)$
  for any $m,n\ge1$.
  On the other hand, we do not know whether the inclusion
  $\QCut(m,n)\subseteq\calE(\nabla\K_{m,n})\cap\RMetPpm(\nabla\K_{m,n})$
  is proper or not.
\end{remark}

Since linear functions can be optimized efficiently over
$\calE(\nabla\K_{m,n})$ by the interior-point method,
Theorem~\ref{theorem:quantum-elliptope2} can be used to give an upper
bound of the maximum quantum violation of any Bell inequality.

For example, Collins and Gisin~\cite{ColGis-JPA04} show that for a
tight Bell inequality for $(3,3)$ settings called
$I_{3322}$~\cite{ColGis-JPA04}:
\[
  f_{3322}=-x_{\X\A_1}-x_{\X\A_2}+x_{\X\B_1}+x_{\X\B_2}
  +x_{\A_1\B_1}+x_{\A_1\B_2}+x_{\A_1\B_3}
  +x_{\A_2\B_1}+x_{\A_2\B_2}-x_{\A_2\B_3}
  +x_{\A_3\B_1}-x_{\A_3\B_2}\le4,
\]
one can achieve $f_{3322}=9/2$ in $\QCut(3,3)$ by using the maximally
entangled state with $\dim\calH_\A=\dim\calH_\B=2$ and appropriate
observables.
Using the SDPA package~\cite{SDPA} for semidefinite programming, we
calculated the maximum of $f_{3322}$ over
$\calE(\nabla\K_{3,3})\cap\RMetPpm(\nabla\K_{3,3})$ as $5.4641$, whose
exact value seems to be $2(\sqrt3+1)$, with unit vectors corresponding
to the nodes $\X,\A_1,\A_2,\A_3,\B_1,\B_2,\B_3$ of $\K_{3,3}$ which lie in a
4-dimensional space and their coordinates in $\RR^4$ are
\begin{align*}
   \vct{w}  =\begin{pmatrix}     1    \\ 0 \\ 0 \\    0     \end{pmatrix},\;
  &\vct{u}_1=\frac{1}{2\sqrt3}
             \begin{pmatrix} 1-\sqrt3 \\ 0 \\ 2 \\ \sqrt3+1 \end{pmatrix},
   \vct{u}_2=\frac{1}{2\sqrt3}
             \begin{pmatrix} 1-\sqrt3 \\ 0 \\-2 \\ \sqrt3+1 \end{pmatrix},
   \vct{u}_3=\begin{pmatrix}     0    \\ 1 \\ 0 \\    0     \end{pmatrix}, \\
  &\vct{v}_1=\frac{1}{2\sqrt3}
             \begin{pmatrix} \sqrt3-1 \\ 2 \\ 0 \\ \sqrt3+1 \end{pmatrix},
   \vct{v}_2=\frac{1}{2\sqrt3}
             \begin{pmatrix} \sqrt3-1 \\-2 \\ 0 \\ \sqrt3+1 \end{pmatrix},
   \vct{v}_3=\begin{pmatrix}     0    \\ 0 \\ 1 \\    0     \end{pmatrix}.
\end{align*}

Like Tsirelson's theorem, Theorem~\ref{theorem:quantum-elliptope2} can
be used to test quantum mechanics itself.
As Froissart pointed out~\cite{Fro-NC81}, finding violation of Bell
inequalities does not prove or disprove quantum mechanics.
However, we do not know if Theorem~\ref{theorem:quantum-elliptope2}
provides a stronger test than Tsirelson's theorem, since we do not
know if there is
$\vct{x}\in\RMetPpm(\nabla\K_{m,n})\setminus\calE(\nabla\K_{m,n})$
such that $\pi(\vct{x})\in\calE(\K_{m,n})$.

Now we consider the set $X_\QB(m,n)$.
It is clear from the definition of $\QCut(m,n)$ that
Theorems~\ref{theorem:nosig-rcmet} and
\ref{theorem:quantum-elliptope2} imply that
$X_\QB(m,n)\subseteq
 \iota(\varphi^{-1}(\calE(\nabla\K_{m,n})))\cap X_\B(m,n)$.
The following theorem replaces the right hand side of this inclusion
with a simpler set.
Here we compare $X_\QB(m,n)$ with the elliptope $\calE(\K_{2m,2n})$ of
the complete bipartite graph $\K_{2m,2n}=(V_{2m,2n},E_{2m,2n})$ as
follows.
We label the nodes of $\K_{2m,2n}$
by $\A_{a,i}$ (for $a\in\{\pm1\}$ and $1\le i\le m$)
and $\B_{b,j}$ (for $b\in\{\pm1\}$ and $1\le j\le n$),
and we identify the $4mn$-dimensional vector space $\RR^{E_{2m,2n}}$
with $\RR^{4mn}$ introduced in Section~\ref{sect:sets} by mapping an
edge $\A_{a,i}\B_{b,j}$ to the coordinate $q_{ab\mid ij}$.

\begin{theorem}
  \label{theorem:quantum-elliptope3}
  $X_\QB(m,n)\subseteq\calE(\K_{2m,2n})\cap X_\B(m,n)$.
\end{theorem}

\begin{proof}
  $X_\QB(m,n)\subseteq X_\B(m,n)$ is obvious.
  $X_\QB(m,n)\subseteq\calE(\K_{2m,2n})$ is immediate from
  Theorem~\ref{theorem:quantum-elliptope2} and the following lemma.
\end{proof}

\begin{lemma}
  \label{lemma:offset-elliptope}
  $\iota(\varphi^{-1}(\calE(\nabla\K_{m,n})))\subseteq\calE(\K_{2m,2n})$.
\end{lemma}

\begin{proof}
  Let $\vct{x}\in\calE(\nabla\K_{m,n})$ and
  $\vct{q}=\iota(\varphi^{-1}(\vct{x}))$.
  We prove $\vct{q}\in\calE(\K_{2m,2n})$.

  By definition of the elliptope $\calE(\nabla\K_{m,n})$,
  there exist unit vectors $\vct{u}_i$ for $1\le i\le m$,
  $\vct{v}_j$ for $1\le j\le n$, and $\vct{w}$ in $\RR^{1+m+n}$
  such that $x_{\X\A_i}=\vct{w}\cdot\vct{u}_i$,
  $x_{\X\B_j}=\vct{w}\cdot\vct{v}_j$ and
  $x_{\A_i\B_j}=\vct{u}_i\cdot\vct{v}_j$.
  By a simple calculation, $\vct{q}$ is written as
  $q_{ab\mid ij}=\vct{u}'_{a,i}\cdot\vct{v}'_{b,j}$
  using vectors $\vct{u}'_{a,i}=(\vct{w}+a\vct{u}_i)/2$ and
  $\vct{v}'_{b,j}=(\vct{w}+b\vct{v}_j)/2$ in $\RR^{1+m+n}$ whose
  lengths are at most $1$.
  By using a well-known technique, we can replace $\vct{u}'_{a,i}$
  and $\vct{v}'_{b,j}$ by unit vectors $\vct{u}''_{a,i}$ and
  $\vct{v}''_{b,j}$ in $\RR^{2m+2n}$ preserving their inner products:
  $q_{ab\mid ij}=\vct{u}''_{a,i}\cdot\vct{v}''_{b,j}$.
  Namely, we add some coordinates to the space to convert the vectors
  $\vct{u}'_{a,i}$ and $\vct{v}'_{b,j}$ to unit vectors in a
  higher-dimensional space, and then we restrict the space to the
  subspace spanned by the $2m+2n$ vectors.
  This proves that $\vct{q}\in\calE(\K_{2m,2n})$.
\end{proof}

However, the usefulness of Theorem~\ref{theorem:quantum-elliptope3} is
yet to be investigated.
For example, Theorem~\ref{theorem:quantum-elliptope3} provides little
information about $X_\QB(3,3)$ since even
$X_\B(3,3)\subsetneq\CutPpm(\K_{6,6})$ holds,
which is proved by enumerating the vertices of $X_\B(3,3)$ by using
cdd~\cite{Fuk:cdd} (cddlib 0.94b).

\section{\Fcf\ inequalities and their tightness}

In this section, we apply results on facet-inducing inequalities
of the cut polytope to the case of $\CutPpm(\K_{m,n})$ to see
the implications of the relationship between the set of classical
correlation functions and the cut polytope $\CutPpm(\K_{m,n})$.

\subsection{Trivial and cycle inequalities as \fcf\ inequalities}
  \label{subsect:trivial-cycle}

Barahona and Mahjoub~\cite{BarMah-MP86} study two classes of
inequalities valid for the cut polytope $\CutPpm(G)$ of an arbitrary
graph $G=(V,E)$, and characterize which of these inequalities are facet
inducing.

For $uv\in E$, \emph{trivial inequalities} are $x_{uv}\le1$ and
$x_{uv}\ge-1$, which are switching equivalent to each other.
They are facet inducing for $\CutPpm(G)$ if and only if the edge $uv$
does not belong to any triangle in $G$.
For a cycle $C=\{u_1u_2,u_2u_3,\dots,u_{l-1}u_l,u_lu_1\}\subseteq E$
and a subset $F\subseteq C$ with $\abs{F}$ odd, a
\emph{cycle inequality} is
$-\sum_{e\in F}x_e+\sum_{e\in C\setminus F}x_e\le\abs{C}-2$.
An example is inequality~(\ref{eq:chsh-cut}), where
$C=\{\A_1\B_1,\A_2\B_1,\A_2\B_2,\A_1\B_2\}$ and $F=\{\A_2\B_2\}$.
Cycle inequalities with a common cycle $C$ and different subsets $F$
are switching equivalent to one another.
They are facet inducing for $\CutPpm(G)$ if and only if the cycle $C$ is
a chordless cycle, i.e.\ no two nodes in $C$ form an edge in $G$ other
than an edge in $C$.
Note that the case of inequality~(\ref{eq:chsh-cut}) satisfies this
condition.

The following theorem follows from this.

\begin{theorem} \label{theorem:trivial-chsh}
  The trivial inequalities
  $x_{\A_i\B_j}\le1$ and $x_{\A_i\B_j}\ge-1$
  for $1\le i\le m$, $1\le j\le n$
  and the CHSH inequalities
  $x_{\A_{i_1}\B_{j_1}}+x_{\A_{i_1}\B_{j_2}}+x_{\A_{i_2}\B_{j_1}}
   -x_{\A_{i_2}\B_{j_2}}\le2$ and
  $x_{\A_{i_1}\B_{j_1}}-x_{\A_{i_1}\B_{j_2}}-x_{\A_{i_2}\B_{j_1}}
   -x_{\A_{i_2}\B_{j_2}}\le2$
  for $1\le i_1,i_2\le m$, $1\le j_1,j_2\le n$, $i_1\ne i_2$,
  $j_1\ne j_2$ are tight \fcf\ inequalities.
\end{theorem}

\subsection{All \fcf\ inequalities with $(4,4)$ settings}

Gisin~\cite{Gis:private-200509} listed all \fcf\
inequalities with a small number of settings per party that involve
small integer coefficients.
As a result, with $(2,2)$ or $(3,s)$ settings with $s=2,3,4$, the CHSH
inequality seems the only nontrivial \fcf\ inequality,
while with $(4,4)$ or more settings other \fcf\
inequalities exist.
He found two facet inducing \fcf\ inequalities with $(4,4)$ settings.

Let
$\sum_{1\le i\le m}\sum_{1\le j\le n}a_{\A_i\B_j}x_{\A_i\B_j}\le a_0$
be a \fcf\ inequality.
We denote this inequality by extracting the coefficients on the
left hand side as
\[
  \left(\begin{array}{cccc}
           &    (\B_1)    & \dots &    (\B_s)    \\
    (\A_1) & a_{\A_1\B_1} & \dots & a_{\A_1\B_s} \\
    \vdots &    \vdots    &       &    \vdots    \\
    (\A_r) & a_{\A_r\B_1} & \dots & a_{\A_r\B_s}
  \end{array}\right)\le a_0.
\]
For example, the CHSH inequality~(\ref{eq:chsh-cut}) is written as
\[
  \left(\begin{array}{ccc}
           & (\B_1) & (\B_2) \\
    (\A_1) &   -1   &   -1   \\
    (\A_2) &   -1   &    1
  \end{array}\right)\le2.
\]

Using this notation, the inequalities found by Gisin are:
\begin{gather}
  \left(\begin{array}{ccccc}
          & (B_1) & (B_2) & (B_3) & (B_4) \\
    (A_1) &  -2   &   2   &   1   &   1   \\
    (A_2) &   1   &   2   &  -2   &  -1   \\
    (A_3) &   1   &   1   &   2   &  -2   \\
    (A_4) &   2   &   1   &   1   &   2
  \end{array}\right)\le10, \label{eq:ineq-4-1} \\
  \left(\begin{array}{ccccc}
          & (B_1) & (B_2) & (B_3) & (B_4) \\
    (A_1) &   2   &   1   &   1   &   0   \\
    (A_2) &   1   &  -1   &  -1   &  -1   \\
    (A_3) &   1   &  -1   &  -1   &   1   \\
    (A_4) &   0   &  -1   &   1   &   0
  \end{array}\right)\le2. \label{eq:ineq-4-2}
\end{gather}
Moreover, Gisin showed that these two are only inequivalent facet
\fcf\ inequalities with $(4,4)$ settings and absolute
values of coefficients at most two.
He also found that there is exactly one facet \fcf\
inequality that involves exactly $(m,n)$ settings with coefficients
$0,\pm1$ for $(m,n)=(4,5),(5,5)$.

To see whether there are any other tight \fcf\ inequalities if we do
not restrict the range of coefficients, we enumerated facet
inequalities of $\CutPpm(\K_{4,4})$ by using cdd~\cite{Fuk:cdd}
(cddlib 0.94b).
After that, we filtered out equivalent \fcf\ inequalities
by using nauty~\cite{Mck:nauty} (nauty 2.2).
We obtained the following result.

\begin{theorem}
  The trivial inequality $x_{\A_1\B_1}\le1$, the CHSH
  inequality~(\ref{eq:chsh-cut}), and the two inequalities
  (\ref{eq:ineq-4-1}) and (\ref{eq:ineq-4-2}) found by Gisin are all
  the inequivalent facet inequalities of $\CutPpm(\K_{4,4})$.
\end{theorem}

\begin{figure}
  \centering
  \includegraphics[width=\textwidth]{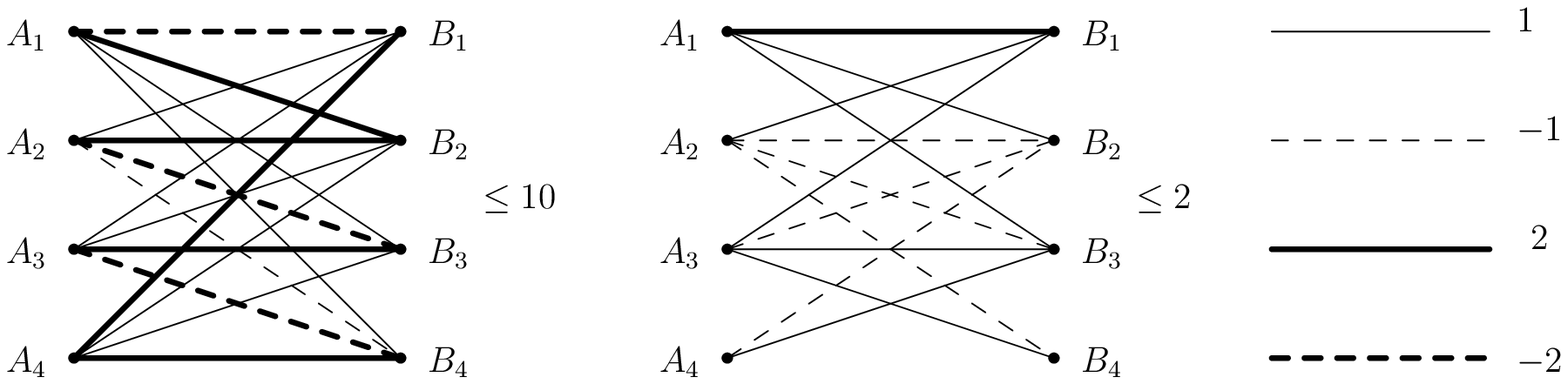}
  \caption{The tight \fcf\ inequalities (\ref{eq:ineq-4-1})
    and (\ref{eq:ineq-4-2}).}
\end{figure}

\subsection{Zero-lifting of \fcf\ inequalities}

If a Bell inequality for $(m,n)$ settings is tight, then it is also a
tight Bell inequality for $(m',n')$, where $1\le m\le m'$ and
$1\le n\le n'$.
This can be proved directly~\cite{Pir-JMP05}, or as a
corollary~\cite{AviImaItoSas-JPA05} of the zero-lifting theorem for
the cut polytope of a general graph~\cite{Des-ORL90}.
Here we prove that the same holds for \fcf\ inequalities.

\begin{theorem}
  \label{theorem:zero-lifting}
  Let $1\le m\le m'$ and $1\le n\le n'$.
  A tight \fcf\ inequality for $(m,n)$ settings is also a tight \fcf\
  inequality for $(m',n')$ settings.
\end{theorem}

\begin{proof}
  It suffices if we prove that an inequality
  $\vct{a}^\trans\vct{x}\le a_0$ which is facet inducing for
  $\CutPpm(\K_{m,n})$ is also facet inducing for
  $\CutPpm(\K_{m',n'})$.

  If the inequality involves only one coordinate, then it is
  necessarily equivalent to the trivial inequality $x_{\A_1\B_1}\le1$,
  and it is facet inducing for $\CutPpm(\K_{m',n'})$ for any $m',n'$
  by Theorem~\ref{theorem:trivial-chsh}.

  If the inequality involves more than one coordinate, then the
  theorem follows by applying the zero-lifting theorem for the cut
  polytope of a general graph~\cite{Des-ORL90} repeatedly, in the same
  way as in the case~\cite{AviImaItoSas-JPA05} of Bell inequalities.
\end{proof}

\subsection{\Fcf\ inequalities for $(3,n)$ settings}

From Gisin's observation that the CHSH inequality is the only
nontrivial \fcf\ inequality for $(2,2)$ or $(3,n)$ settings for
$n=2,3,4$, one may conjecture that this is true for general $n$.
Here we prove this.

\begin{theorem}
  If $\min\{m,n\}\le3$, then the inequalities in
  Theorem~\ref{theorem:trivial-chsh} are all the tight
  \fcf\ inequalities for $(m,n)$ settings.
\end{theorem}

\begin{proof}
  If $\min\{m,n\}\le3$, then $\K_{m,n}$ is not contractible to $\K_5$.
  Barahona and Mahjoub~\cite{BarMah-MP86} show that if a graph $G$ is
  not contractible to $\K_5$, then all facet-inducing inequalities of
  $\CutPpm(G)$ are either the trivial or the cycle inequality.
  The only facet-inducing cycle inequality of $\CutPpm(\K_{m,n})$ is the
  CHSH inequality since the only chordless cycle in a complete
  bipartite graph is a cycle of length four.
\end{proof}

A similar result was shown in the context of Bell inequalities by Collins
and Gisin~\cite{ColGis-JPA04}: a nonnegativity inequality
$p_{\A_i\B_j}\ge0$ and the CHSH inequality are all the inequivalent
Bell inequalities for $(m,n)$ settings with $\min\{m,n\}=2$.
As is pointed out in \cite{AviImaItoSas-JPA05}, this can also be
proved using Barahona and Mahjoub's result.

\subsection{Triangular elimination and \fcf\ inequalities}

\emph{Triangular elimination}~\cite{AviImaItoSas-JPA05} is a method to
convert inequalities valid for the cut polytope $\CutPpm(\K_N)$ of the
complete graph to inequalities valid for
$\CutPpm(\nabla\K_{m,n})$, which correspond to Bell inequalities via the
covariance mapping, preserving their facet-inducing property.
In \cite{AviImaIto:trielim}, this result is extended
to the case of general graphs and
in particular the case of bipartite graphs.
The inequalities constructed by triangular elimination in the case
of bipartite graphs, can be regarded as \fcf\ inequalities,
as shown by
Proposition~\ref{prop:cf-kmn}~(i).

We describe triangular elimination from $\CutPpm(\K_N)$ to
$\CutPpm(\K_{m,n})$ by example.
A complete definition and a proof of relevant theorems are stated in
\cite{AviImaIto:trielim}.

The inequality
\begin{equation}
  -x_{\A_1\A_2}-x_{\A_1\A_3}-x_{\A_2\A_3}-x_{\B_1\B_2}
    +\sum_{i=1,2,3}\sum_{j=1,2}x_{\A_i\B_j}\le2
  \label{eq:pentagonal}
\end{equation}
is facet inducing for $\CutPpm(\K_5)$.
It is known as the \emph{pentagonal inequality} and is a special case of
a hypermetric inequality~\cite[Chapter~28]{DezLau:cut97}.
This inequality is not a \fcf\ inequality because it depends on the
coordinates $x_{\A_1\A_2},x_{\A_1\A_3},x_{\A_2\A_3},x_{\B_1\B_2}$,
which cannot appear in a \fcf\ inequality.
We eliminate a coordinate $x_{\A_1\A_2}$ by appending a new node,
which we label $\B_{12}$, and adding a triangle inequality.
The \emph{triangle inequality}~\cite[Chapter~27]{DezLau:cut97} is an
inequality in the form $-x_{uv}-x_{uw}-x_{vw}\le1$ or in the form
$-x_{uv}+x_{uw}+x_{vw}\le1$, and also facet inducing for the cut
polytope of the complete graph.
In this case, adding a triangle inequality
$x_{\A_1\A_2}-x_{\A_1\B_{12}}+x_{\A_2\B_{12}}\le1$ eliminates the
coordinate $x_{\A_1\A_2}$.
Similarly, we append three more nodes $\B_{13},\B_{23},\A_{12}$ and
add three appropriate triangle inequalities to eliminate
$x_{\A_1\A_3},x_{\A_2\A_3},x_{\B_1\B_2}$.
This gives the inequality
\begin{equation}
  -x_{\A_1\B_{12}}+x_{\A_2\B_{12}}
  -x_{\A_1\B_{13}}+x_{\A_3\B_{13}}
  -x_{\A_2\B_{23}}+x_{\A_3\B_{23}}
  -x_{\A_{12}\B_1}+x_{\A_{12}\B_2}
  +\sum_{i=1,2,3}\sum_{j=1,2}x_{\A_i\B_j}\le6,
  \label{eq:pentagonal-trielim}
\end{equation}
or
\[
  \left(\begin{array}{cccccc}
             & (B_1) & (B_2) & (B_{12}) & (B_{13}) & (B_{23}) \\
    (A_1)    &   1   &   1   &    -1    &    -1    &     0    \\
    (A_2)    &   1   &   1   &     1    &     0    &    -1    \\
    (A_3)    &   1   &   1   &     0    &     1    &     1    \\
    (A_{12}) &  -1   &   1   &     0    &     0    &     0
  \end{array}\right)\le6.
\]
It is proved in~\cite{AviImaIto:trielim} that the inequality constructed
in this way is facet inducing for the cut polytope of the complete
bipartite graph ($\CutPpm(\K_{4,5})$ in this case).

Tight \fcf\ inequalities constructed in this way can be seen as
special cases of Bell inequalities constructed by triangular
elimination from $\K_N$ to $\nabla\K_{m,n}$ which happen to be
\fcf\ inequalities.
For example, (\ref{eq:pentagonal}) is also facet inducing for
$\CutPpm(\K_6)$ with nodes $\A_1,\A_2,\A_3,\B_1,\B_2,\X$ because of the
zero-lifting theorem.
Applying triangular elimination from $\CutPpm(\K_6)$ to
$\CutPpm(\K_{1,4,5})$, we construct the same
inequality~(\ref{eq:pentagonal-trielim}).

\subsubsection{Counting \fcf\ inequalities constructed by triangular elimination}

Table~\ref{table:count} shows the number of tight \fcf\
inequalities obtained by triangular elimination from facet
inequalities of $\CutPpm(\K_N)$ for $N\le9$.
The lists of facet inequalities of $\CutPpm(\K_N)$ are obtained from
\cite{SMAPO}.

\begin{table}[h]
  \centering
  \caption{\label{table:count}
    The number of facets of $\CutPpm(\K_N)$,
    the number of tight Bell inequalities
    obtained as the triangular eliminations of the facets of
    $\CutPpm(\K_N)$,
    and the number of tight \fcf\ inequalities in them.
    Two facets which can be transformed by permutation or switching are
    considered identical, and two equivalent Bell or \fcf\
    inequalities are considered identical.
    An asterisk (*) indicates the value is a lower bound.
    The lists of facet inequalities of $\CutPpm(\K_N)$ are obtained from
    \cite{SMAPO}.
    The number of Bell inequalities is taken from
    \cite{AviImaItoSas-JPA05}.}
  \catcode`\@\active \def@{\phantom{,}}
  \catcode`\?\active \def?{\phantom{0}}
  \catcode`\=\active \def={\phantom{*}}
  \begin{tabular}{cccc} \hline
    $N$ & Facets of $\CutPpm(\K_N)$ &
                     Tight Bell ineqs. &
                     Tight \fcf\ ineqs. \\ \hline
     3  & ???@??1= & ???@???@??2= & ??@???@??1= \\
     4  & ???@??1= & ???@???@??2= & ??@???@??1= \\
     5  & ???@??2= & ???@???@??8= & ??@???@??4= \\
     6  & ???@??3= & ???@???@?22= & ??@???@?10= \\
     7  & ???@?11= & ???@???@323= & ??@???@107= \\
     8  & ???@147* & ???@?40,399* & ??@??9,159* \\
     9  & 164,506* & 201,374,783* & 37,346,094* \\ \hline
  \end{tabular}
\end{table}

\subsubsection{A family of \fcf\ inequalities constructed by triangular elimination}

The following theorem follows from the family of Bell
inequalities~\cite{AviImaItoSas-JPA05} constructed from hypermetric
inequalities~\cite[Section~28]{DezLau:cut97} valid for the cut
polytope of the complete graph.

\begin{theorem} \label{theorem:facet-kmn-hm}
  Let
  $b_{\A_1},\dots,b_{\A_s},\allowbreak b_{\B_1},\dots,b_{\B_t}$
  be integers such that
  $\sum_{i=1}^s b_{\A_i}+\sum_{j=1}^t b_{\B_j}=1$.
  Then,
  \begin{enumerate}[(i)]
  \item
    The inequality
    \begin{multline}
       \sum_{1\le i\le s}\sum_{1\le j\le t}
         b_{\A_i}b_{\B_j} x_{\A_i\B_j} \\
      +\sum_{1\le i<i'\le s}
         (b_{\A_i}b_{\A_{i'}} x_{\A_i\B_{ii'}}
          -\abs{b_{\A_i}b_{\A_{i'}}} x_{\A_{i'}\B_{ii'}})
      +\sum_{1\le j<j'\le t}
         (b_{\B_j}b_{\B_{j'}} x_{\A_{jj'}\B_j}
          -\abs{b_{\B_j}b_{\B_{j'}}} x_{\A_{jj'}\B_{j'}}) \\
      \ge
       \sum_{1\le i\le s}\sum_{1\le j\le t}
         b_{\A_i}b_{\B_j}
      +2\sum_{\substack{1\le i<i'\le s \\ b_{\A_i}b_{\A_{i'}}<0}}
         b_{\A_i}b_{\A_{i'}}
      +2\sum_{\substack{1\le j<j'\le t \\ b_{\B_j}b_{\B_{j'}}<0}}
         b_{\B_j}b_{\B_{j'}}
      \label{eq:hm-kmn}
    \end{multline}
    is a valid \fcf\ inequality.
  \item
    The \fcf\ inequality~(\ref{eq:hm-kmn}) is facet inducing if one of
    the following conditions is satisfied.
    \begin{enumerate}[(a)]
    \item
      For some $l>1$, the integers
      $b_{\A_1},\dots,b_{\A_s},\allowbreak b_{\B_1},\dots,b_{\B_t}$
      contain $l+1$ entries equal to $1$ and $l$ entries equal to
      $-1$, and the other entries (if any) are equal to $0$.
    \item
      At least $3$ and at most $n-3$ entries in
      $b_{\A_1},\dots,b_{\A_s},\allowbreak b_{\B_1},\dots,b_{\B_t}$
      are positive, and all the other entries are equal to $-1$.
    \end{enumerate}
  \end{enumerate}
\end{theorem}

\begin{proof}
  \begin{enumerate}[(i)]
  \item
    If we let $b_\X=0$ in Theorem~3.1 of \cite{AviImaItoSas-JPA05} and
    rewrite the resulting Bell inequality in variables in
    $\vct{x}\in\RR^{\nabla E_{m,n}}$ by using the covariance mapping, we
    obtain the inequality~(\ref{eq:hm-kmn}).
    What we obtain is an inequality valid for
    $\CutPpm(\nabla\K_{m,n})$, but it is also an inequality valid for
    $\CutPpm(\K_{m,n})$ since it does not contain any variables
    related to the node $\X$.
  \item
    By Theorem~3.1~(ii) of \cite{AviImaItoSas-JPA05},
    the inequality~(\ref{eq:hm-kmn}) is facet-inducing for
    $\CutPpm(\nabla\K_{m,n})$.
    From a well-known fact in the theory of polytopes that projecting
    out unused terms preserves facet-inducing inequalities (see
    e.g.\ Lemma~26.5.2~(ii) of \cite{DezLau:cut97}),
    the inequality~(\ref{eq:hm-kmn}) is facet-inducing also for
    $\CutPpm(\K_{m,n})$.
  \end{enumerate}
\end{proof}

\section{Concluding remarks}
  \label{sect:conclusion}

We conclude the paper with some open problems.
The projection of $\QCut(m,n)$ to $E_{m,n}$ is described directly
in terms of the elliptope, but how is the set $\QCut(m,n)$ described?
In other words, how close are the sets $\QCut(m,n)$ and
$\calE(\nabla\K_{m,n})\cap\RMetPpm(\nabla\K_{m,n})$ ?

The upper bound $5.4641$ of $f_{3322}$ in quantum correlation
experiments obtained by using
Theorem~\ref{theorem:quantum-elliptope2} differs from the known lower
bound $9/2$, which is achievable in the two-qubit system.
If the upper bound can be improved, it may lead to a refinement of
Theorem~\ref{theorem:quantum-elliptope2}.
It may be the case that the lower bound $9/2$ is the maximum in the
two-qubit system but not in a quantum system with a higher
dimension, given that the two-qubit and two-qutrit (three-level)
systems are quite different in terms of the ``strength''
(\emph{relevance}~\cite{ColGis-JPA04}) of the CHSH
inequalities~\cite{ItoImaAvi-PRA06}.

As we stated in Section~\ref{sect:cor-func},
Tsirelson~\cite{Tsi-JSovM87} gives an upper bound on the quantum
violation of any \fcf\ inequality by using Grothendieck's
inequality~\cite{Kri-AdvMath79}.
If Grothendieck's inequality can be extended to $\nabla\K_{m,n}$,
we can combine it
with Theorem~\ref{theorem:quantum-elliptope2} to obtain an upper bound
of the quantum violation of any Bell inequalities.

Gill~\cite{Gil-open26} asks whether there exists a tight Bell
inequality holding for all quantum correlation experiments other than
the trivial ones representing nonnegativity of probabilities.
Such an inequality corresponds to a facet of $X_\HDB(m,n)$ which is
valid for $X_\QB(m,n)$ but not for $X_\B(m,n)$.
Asking the same question for tight \fcf\ inequalities corresponds to the
question of whether the elliptope $\calE(\K_{m,n})$ has a facet in common
with $\CutPpm(\K_{m,n})$ other than the trivial ones.
The facial structure of the elliptope of the complete graph is studied
by Laurent and Poljak~\cite{LauPol-LAA95}, but we are not aware of any
similar results for the bipartite graph.

In connection to the relation between Bell inequalities and quantum
games explored by Cleve, H{\o}yer, Toner and
Watrous~\cite{CleHoyTonWat-CCC04}, \fcf\ inequalities
correspond to XOR games.
They pointed out that from Tsirelson's theorem, the winning
probability of XOR games by quantum players can be computed
efficiently by
using semidefinite programming and gave an upper bound on the quantum
winning probability by using Grothendieck's inequality.
In a similar way, Theorem~\ref{theorem:quantum-elliptope2} gives an
efficient way to compute an upper bound of the quantum winning
probability of general binary games, and if Grothendieck's inequality
can be generalized to $\nabla\K_{m,n}$, an analytical upper bound will
be given also for binary games.

Extending individual inequalities such as (\ref{eq:ineq-4-1}) and
(\ref{eq:ineq-4-2}) to classes of inequalities like Collins and
Gisin~\cite{ColGis-JPA04} did by introducing the
$I_{mmvv}$
inequalities, and researchers in polyhedral combinatorics have done
for many classes of inequalities for the cut polytope of the complete
graph~\cite[Chapters~27--30]{DezLau:cut97}, will give better
understanding of these \fcf\ inequalities.
In addition, it would be useful if we have an analytical bound on the
maximum quantum violation for families of
\fcf\ inequalities.

\section*{Acknowledgment}

The authors would like to thank Nicolas Gisin for discussions at the
ERATO Conference on Quantum Information Science (EQIS2005), Aug.\
26--30, 2005, Tokyo, where he suggested we turn our attention to \fcf\
inequalities.
Research of the first author is supported by N.S.E.R.C.\ and F.Q.R.N.T.,
and the third author is grateful for a support by the
Grant-in-Aid for JSPS Fellows.

\bibliography{bell}

\appendix

\section{\Fcf\ inequalities in $(4,5)$ settings}

We enumerated facet inequalities of $\CutPpm(\K_{4,5})$
by using cdd~\cite{Fuk:cdd} (cddlib version 0.94b).%
\footnote{We used the cut cone $\CUT(\K_{4,5})$ instead of the cut
  polytope $\CutPpm(\K_{4,5})$ to reduce the number of facets.
  This does not essentially change the output since any facet of
  $\CutPpm(\K_{4,5})$ has a corresponding facet of $\CUT(\K_{4,5})$
  which is switching equivalent to it
  (see e.g.\ \cite[Section~26.3.2]{DezLau:cut97}).}
The computation was aborted (seemingly because it ran out the memory),
but the partial result shows some of the facets of $\CutPpm(\K_{4,5})$
that are not the zero-lifting of any facets of $\CutPpm(\K_{4,4})$:
\begin{gather*}
  % (48) in k45-cone.idx
  \left(\begin{array}{cccccc}
          & (B_1) & (B_2) & (B_3) & (B_4) & (B_5) \\
    (A_1) &   1   &   0   &   0   &   0   &   1   \\
    (A_2) &   1   &   1   &   1   &   0   &  -1   \\
    (A_3) &   1   &   0   &  -1   &   1   &  -1   \\
    (A_4) &  -1   &   1   &   0   &   1   &   1
  \end{array}\right)\le6, \\
  % (138) in k45-cone.idx
  \left(\begin{array}{cccccc}
          & (B_1) & (B_2) & (B_3) & (B_4) & (B_5) \\
    (A_1) &   2   &   1   &   1   &   1   &   1   \\
    (A_2) &   0   &   1   &  -1   &   1   &  -1   \\
    (A_3) &   0   &  -1   &   1   &   1   &  -1   \\
    (A_4) &  -2   &   1   &   1   &   1   &   1
  \end{array}\right)\le8, \\
  % (2298) in k45-cone.idx
  \left(\begin{array}{cccccc}
          & (B_1) & (B_2) & (B_3) & (B_4) & (B_5) \\
    (A_1) &   2   &   1   &   1   &   1   &   1   \\
    (A_2) &  -1   &   1   &   2   &   1   &  -1   \\
    (A_3) &  -1   &   2   &   1   &  -1   &   1   \\
    (A_4) &   0   &   2   &  -2   &   1   &  -1
  \end{array}\right)\le10, \\
  % (3105) in k45-cone.idx
  \left(\begin{array}{cccccc}
          & (B_1) & (B_2) & (B_3) & (B_4) & (B_5) \\
    (A_1) &   1   &   2   &   1   &   1   &  -1   \\
    (A_2) &   0   &   2   &  -1   &  -1   &   2   \\
    (A_3) &   1   &  -1   &   1   &  -2   &   1   \\
    (A_4) &   0   &  -1   &   1   &   2   &   2
  \end{array}\right)\le10.
\end{gather*}

\end{document}